\documentclass[11pt,conference,letterpaper,twocolumn]{IEEEtran}

\usepackage[english]{babel}    
\usepackage[latin1]{inputenc} 
\usepackage{float} 
\usepackage{graphicx} 
\usepackage{url} 
\usepackage{multirow} 
\usepackage{amssymb}
\usepackage{array,color}
\usepackage[array,color]{colortbl}
\graphicspath{{./graphics/}}

\newtheorem{defi}{Definition}
\newtheorem{prop}{Proposition}

\newenvironment{proof}{\begin{trivlist}\item[]{\em Proof.\\ }}%
{\samepage \hfill$\Box$ \\ \end{trivlist}}

\makeatletter
\def\hlinewd#1{%
\noalign{\ifnum0=`}\fi\hrule \@height #1 %
\futurelet\reserved@a\@xhline}
\makeatother

\title{Malware Detection using Attribute-Automata to parse Abstract Behavioral Descriptions}

\author{Grégoire Jacob$^{(1/2)}$, Hervé Debar$^{(1)}$, Eric Filiol$^{(2)}$\\[0.6em]
\small $^1$ France Télécom R\&D, Caen, France\\
\small \texttt{\{gregoire.jacob|herve.debar\}@orange-ftgroup.com}\\[0.6em]
\small $^2$ ESIEA,\\ \small Operational Virology and Cryptology Lab., Laval ou Paris, France\\
\small \texttt{eric.filiol@esat.terre.defense.gouv.fr}
}

\begin{document}

\maketitle

\begin{abstract}
Most behavioral detectors of malware remain specific to a given language and platform, mostly PE executables for Windows. The objective of this paper is to define a generic approach for behavioral detection based on two layers respectively responsible for abstraction and detection. The first abstraction layer remains specific to a platform and a language. This first layer interprets the collected instructions, API calls and arguments and classifies these operations as well as the involved objects according to their purpose in the malware lifecycle. The second detection layer remains generic and is totally interoperable between the different abstraction components. This layer relies on parallel automata parsing attribute-grammars where semantic rules are used for object typing (object classification) and object binding (data-flow). To feed detection and to experiment with our approach we have developed two different abstraction components: one processing system call traces from native code and one processing the VBScript interpreted language. The different experimentations have provided promising detection rates, in particular for script files (89\%), with almost none false positives. In the case of process traces, the detection rate remains significant (51\%) but could be increased by more sophisticated collection tools.\\\\[-6pt]{} 
\end{abstract}

%\tableofcontents
\pagenumbering{arabic} 

\section{Introduction}
	Malware behavioral detection is an active research field since the behavioral approach should theoretically be able to detect, if not innovative malware, at least unknown malware reusing variations of known techniques. With regards to actual known methods of behavioral detection, most of them rely on specific characteristics; this enables evasion through simple functional modifications: the regular reappearance of new versions from known strains, de facto multiplying the detection signatures, is an obvious consequence. The leading objective of this article is to provide a generic grammar for global malicious behaviors in order to build efficient and resilient detection automata. Genericity is introduced in the process by an abstraction from the platform and the language.\\ 
\indent The use of deterministic finite automata to detect behaviors is an attractive idea since the complexity of these algorithms remains linear and thus acceptable for operational deployment. Published in 1995, \cite{CM95} uses automata to structure the possible sequences of operations making up behaviors, but this, without real control of the data flow. Since, various works in computer security have focused on this notion of data flow, finally leading to the apparition of tainting techniques to detect malicious uses of data \cite{NS05}. These techniques have exhibited significant successes and the notion of data flow control is now broadly used, like in intrusion detection \cite{BCS06} and malware behavior extraction \cite{CJK07}. These two articles use automata to model different sequences of system calls constituting respectively attacks and malicious behaviors. The data flow is then captured by analysis of the parameters collected along these system calls. In parallel, a similar detection technique focusing on self-reproduction behaviors has been published, the authors arguing that self-replication was the only decisive characteristic common to all malware \cite{MCD08}. Our approach also combines these two aspects: detection by automata and flow control. However, it does not restrict itself to a single behavior. We think that monitoring a unique behavior without correlation with other potential malicious actions may prove insufficient for reliable detection. We have thus defined a generative model to describe several classes of malicious behaviors and assessed the detection of different behaviors descriptions.\\ % in order to assess their significance. \\
\indent In fact, the detection method we use can be related to methods of attack scenario recognition in intrusion detection \cite{AZ08}. In these methods, isolated alerts are correlated into a scenario by parsing attribute-grammars annoted with semantic rules to guarantee the flow between the different related alerts. In the context of malware detection, malicious behaviors are also described through attribute-grammars: on the one hand, the syntactic rules describe the possible combinations for the basic operations making up the behavior, on the other hand, the semantic rules control the data flow between the elements involved in these operations, but also associate to these elements a potential purpose in the malware lifecycle (installation, communication or execution). The detection process is therefore achieved by parsing techniques. In order to feed detection, an additional abstraction process is required to translate the observable data into the model. Abstraction provides a generic level of description where the processed data become detached from the specificities of the platform and any programming language. With regards to the generation of the grammatical behavior descriptions, the identification of the system objects with a potential use for malware or the language abstraction: all these operations require a first configuration step as described in Figure 1. However, contrary to other detection methods, the configuration focuses both on critical system objects, which remain enumerable in a standard applicative environment, and innovative malware, which are scarce among the numerous variants of known malware.  This contribution can finally be sum up in three main points:
\begin{itemize}
\item A model of malicious behaviors by attribute-grammars with semantic rules for object binding in order (control the data flow) and object typing (convey the potential purpose of objects in the malware lifecycle).
\item The introduction of an abstraction layer for translation into the model, detaching the detection process from the specificities of the platform and programming languages. As proofs of concept, two specific abstraction components have been developed for the analysis of executable traces and interpreted scripts.
\item Several generic automata to detect different classes of behaviors. Different assessments are made both from the theoretical perspectives (time and space complexities) and from the operational perspectives (coverage and performance).
\end{itemize}

\begin{figure*}[!ht]
  \centering
  \includegraphics*[scale=0.45]{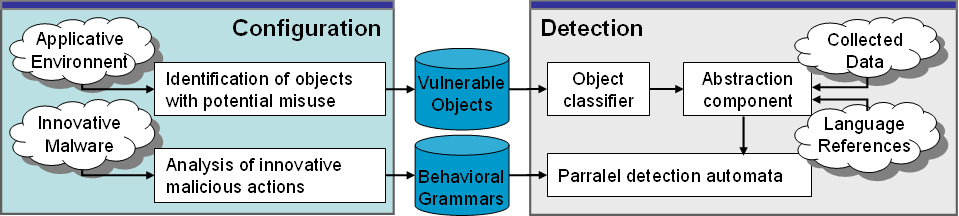} 
  \label{fig:confdet}
  %\usefont{OT1}{cmr}{b}{n}
  \caption{Configuration and detection processes.}
\end{figure*}

\indent The article is articulated as follows. For the article to be self-contained, it first introduces in Section II a behavioral model based on attribute-grammars which recalls and synthesizes results from previous works. Section III deals with the abstraction process from the collected data to the behavioral model. Section IV describes the detection process based on parsing automata. An implementation is presented in Section V whose results are interpreted and commented in Section VI.

\section{Formalization of malicious behaviors using attribute-grammars}

	From a theoretical perspective an attribute-grammar is a Context-Free Grammar (CFG) enriched with semantic attributes and rules \cite{KN68}. A complete definition is detailed below. In the particular case of behavioral detection, each start symbols begins the description of a new malicious behavior. The production rules then describe the different technical solutions to achieve this behavior whereas the terminal symbols of the grammar correspond to the data collected through the abstraction layer (interpretations of instructions, API calls, arguments) \cite{JFD07,JFD08}. Based on these principles, the coming sections more specifically address the grammatical model of the behavior descriptions.\\

\begin{defi} An attributed-grammar $G_A$ is a triplet $<$$G,D,E$$>$ where:\\
- $G$ is originally a context-free grammar $<$$V,\Sigma ,S,P$$>$,\\
- let $Att=Syn \uplus Inh$ be a set of attributes divided between the synthesized and the inherited attributes, and $D=\cup_{\alpha\in Att} D_\alpha$ be the union of their sets of values,\\
- let $att:X \in \{V \cup \Sigma \} \longrightarrow att(X)\in Att^{*}$ be an attribute assignment function,\\
- every production rule $\pi \in P$ of the form $Y_0\longrightarrow Y_1...Y_n$ determines a set of attributes $Var_{\pi}= \cup_{i\in\{0,...,n\}}\{Y_i.\alpha \mid \alpha \in att(Y_i)\}$ partitioned between inner variables:
$In_{\pi}=\{Y_0.\alpha \mid \alpha \in att(Y_0)\cap Syn\}\cup \{Y_i.\alpha \mid i\not = 0,\alpha \in att(Y_i)\cap Inh\}$,\\
and outer variables:
$Out_{\pi}=Var_{\pi}\setminus In_{\pi}$,\\
- $E$ is a set of semantic rules such as for any production rule $\pi\in P$, for each inner variable $Y_i.\alpha \in In_{\pi}$, there is exactly one rule of the form $Y_i.\alpha = f(Y_1.\alpha_1...Y_n.\alpha_n)$ where $Y_j.\alpha_k \in Out_{\pi}$ and $f:D_{\alpha_1}\times ...\times D_{\alpha_n} \rightarrow D_{\alpha}$.\\
\end{defi}

	\subsection{Malicious Behavior Language}
	A generic programming language is required to describe any malicious behavior: to this purpose, a generic language, called the Malicious Behavior Language (MBL), has been developed in a previous article giving its syntax and operational semantic \cite{JFD07}. Most malicious behaviors can be described by sub-grammars of this generative grammar, that is to say: the different behavior languages are included in the MBL. In a few word, the MBL relies on basic arithmetic and control operations guaranteeing the Turing completeness as well as additional interaction operations: commands (open, create, close, delete) or inputs/outputs (send, receive, signal, wait). As discussed in the article, malware being highly resilient and adaptable by nature, interaction operations are key features of this language.\\
\indent In addition to the syntax of these operations, a type system has been provided for the external objects adressed by the interactions. The different objects have been typed according to their potential use in the malware lifecycle: permanent objects ($obj\_perm$), temporary objects ($obj\_temp$), booting object ($obj\_boot$), communicating objects ($obj\_com$), self-reference ($this$). A partial order has been defined on these types according to their subset inclusion, as shown in Figure 2. The semantic attributes enriching attribute-grammars will prove useful to deploy such a type system. In fact, semantic attributes and rules can be used for several purposes:
\begin{description}
\item[\textbf{Object binding:}]  $\qquad\qquad\;$ This mechanism identifies the different instances of objects and variables and guarantees they are coherently used. Object binding can simply be achieved by affecting specific attributes called identifiers to the terminal symbols representing these objects (denoted $objId$ in the semantic rules). In the context of interactions, object binding is also used for studying the data-flow between objects. As shown in the next section, it is particularly important in behaviors such as duplication where the viral code is transferred from the self-reference to a target object.
\item[\textbf{Object typing:}] $\qquad\quad\;\;$ A type attribute can also be affected  to a given object (denoted $objTp$). Types are attached to objects according to their potential uses. This is particularly important to discover malicious purposes such as booting objects in the case of residency or communicating objects in the case of propagation. Additional characterisation of the objects can be achieved through additional attributes. For example, an attribute can be defined to store the nature of an object (denoted $objNat$): variable, file, registry key, network socket, mail, etc. Typing may then be refined according to these additional attributes.
\end{description}
	
\begin{figure}[!ht]
  \centering
  \includegraphics*[scale=0.34]{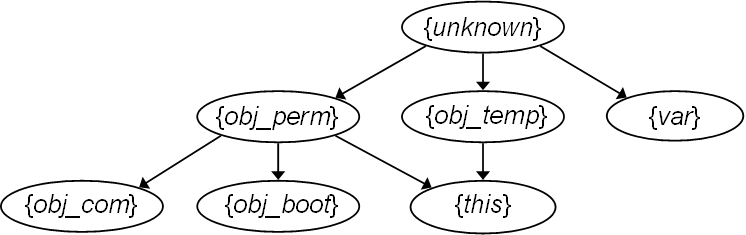} 
  \label{fig:typeposet}
  %\usefont{OT1}{cmr}{b}{n}
  \caption{Partial order on object types. %\usefont{OT1}{cmr}{m}{n} 
The partial order has been defined according the inclusions of the different subsets which are represented by the different edges of the Hasse Diagram (for example: $obj\_perm\leq obj\_boot$). In fact, the set inclusions correspond to a specialization of the objects according to their use in the malware lifecycle.}
\end{figure}

	\subsection{Descriptions of malicious behaviors}
	Use of this grammar is best illustrated by different examples of behavioral descriptions. In this article, we focus ourselves on four different behaviors. Because their whole descriptions would be too tedious, this section only covers two of the most spread behaviors: duplication and propagation. The other considered behaviors are residency (automatic start of the malware) and overinfection tests. Their preliminary descriptions, as well as others' (infection, mutation, activity tests), have been manually generated in the original paper by analysing a large pool of malware \cite{JFD07}. Since the behavioral descriptions only convey the most generic features of the malicious behaviors, manual generation of these behavioral signatures can be considered more easily than for the binary signatures used in scanning techniques.\\

\subsubsection{Duplication}
	Duplication is achieved by copying data from the self-reference towards a newly created permanent object. It can be described by the following syntactic production rules (grey) and their additional semantic rules (white). The different syntactic derivations correspond to the different duplication techniques: single-block read/write operations, interleaved read/write and direct copy, with possible operation permutations. The semantic rules are more interesting because they can both guarantee the data-flow between objects: the read and write operations must refer to the same variable (Object Binding: $<$$Write$$>$.varId$\,=\,<$$Read$$>$.varId), and guarantee the maliciousness of the behavior: the open and read operations must refer to the self-reference to be a real duplication (Object Typing: $<$$Duplicate$$>$.srcTp$\,=\,this$).\\[1em]	
\begin{scriptsize}
\noindent 
\begin{tabular}{@{}l@{$\;\,$ }l@{ }l@{ }l}
\hline
\hline
\rowcolor[gray]{.9} $(i)$ & $<$$Duplicate$$>$ & $::=$ & $<$$Create$$><$$Open$$><$$Read$$><$$Write$$>$\\
\rowcolor[gray]{.9} & & $\; \mid $ & $<$$Open$$><$$Create$$><$$Read$$><$$Write$$>$ \\
\rowcolor[gray]{.9} & & $\; \mid $ & $<$$Open$$><$$Read$$><$$Create$$><$$Write$$>$ \\
\rowcolor[gray]{.9} & & $\; \mid $ & $<$$Open$$><$$Create$$><$$IntrlvRW$$>$ \\
\rowcolor[gray]{.9} & & $\; \mid $ & $<$$Create$$><$$Open$$><$$IntrlvRW$$>$ \\
\multicolumn{2}{@{}l}{$\{\;<$$Duplicate$$>$.srcId} & $\, =$ & $<$$Open$$>$.objId \\
\multicolumn{2}{@{}l}{$\;\;\,<$$Read$$>$.objId} & $\, =$ &  $<$$Duplicate$$>$.srcId \\
\multicolumn{2}{@{}l}{$\;\;\,<$$IntrlvRW$$>$.obj1Id} & $\, =$ &  $<$$Duplicate$$>$.srcId \\ 
\multicolumn{2}{@{}l}{$\;\;\,<$$Duplicate$$>$.srcTp} & $\, =$ & $this$ \\
\multicolumn{2}{@{}l}{$\;\;\,<$$Open$$>$.objTp} & $\, =$ & $<$$Duplicate$$>$.srcTp \\
\multicolumn{2}{@{}l}{$\;\;\,<$$Read$$>$.objTp} & $\, =$ & $<$$Duplicate$$>$.srcTp \\
\multicolumn{2}{@{}l}{$\;\;\,<$$IntrlvRW$$>$.obj1Tp} & $\, =$ & $<$$Duplicate$$>$.srcTp \\ 
\multicolumn{2}{@{}l}{$\;\;\,<$$Duplicate$$>$.targId} & $\, =$ & $<$$Create$$>$.objId \\
\multicolumn{2}{@{}l}{$\;\;\,<$$Write$$>$.objId} & $\, =$ & $<$$Duplicate$$>$.targId \\
\multicolumn{2}{@{}l}{$\;\;\,<$$IntrlvRW$$>$.obj2Id} & $\, =$ &  $<$$Duplicate$$>$.targId \\ 
\multicolumn{2}{@{}l}{$\;\;\,<$$Duplicate$$>$.targTp} & $\, =$ & $obj\_perm$ \\
\multicolumn{2}{@{}l}{$\;\;\,<$$Create$$>$.objTp} & $\, =$ & $<$$Duplicate$$>$.targTp  \\
\multicolumn{2}{@{}l}{$\;\;\,<$$Write$$>$.objTp} & $\, =$ & $<$$Duplicate$$>$.targTp  \\ 
\multicolumn{2}{@{}l}{$\;\;\,<$$IntrlvRW$$>$.obj2Tp} & $\, =$ &  $<$$Duplicate$$>$.targTp \\ 
\multicolumn{2}{@{}l}{$\;\;\,<$$Write$$>$.varId} & $\, =$ & $<$$Read$$>$.varId $\qquad \qquad \quad\}$\\
\rowcolor[gray]{.9} & & $\; \mid $ & $<$$DrctCopy$$>$ \\
\multicolumn{2}{@{}l}{$\{\;<$$Duplication$$>$.srcId} & $\, =$ & $<$$DrctCopy$$>$.obj1Id \\
\multicolumn{2}{@{}l}{$\;\;\,<$$Duplication$$>$.srcTp} & $\, =$ & $this$ \\
\multicolumn{2}{@{}l}{$\;\;\,<$$DrctCopy$$>$.obj1Tp} & $\, =$ & $<$$Duplicate$$>$.srcTp \\
\multicolumn{2}{@{}l}{$\;\;\,<$$Duplicate$$>$.targId} & $\, =$ & $<$$DrctCopy$$>$.obj2Id \\
\multicolumn{2}{@{}l}{$\;\;\,<$$Duplicate$$>$.targTp} & $\, =$ & $obj\_perm$ \\
\multicolumn{2}{@{}l}{$\;\;\,<$$DrctCopy$$>$.obj2Tp} & $\, =$ & $<$$Duplicate$$>$.targTp $\qquad \}$\\
\rowcolor[gray]{.9}  $(ii)$ & $<$$Create$$>$ & $::=$ & $create \quad object;$ \\
\multicolumn{2}{@{}l}{$\{\;<$$Create$$>$.objId} & $\, =$ &$object$.objId \\
\multicolumn{2}{@{}l}{$\;\;\,object$.objTp} & $\, =$ & $<$$Create$$>$.objTp $\qquad \quad \;\; \}$\\
\rowcolor[gray]{.9} $(iii)$ & $<$$Open$$>$ & $::=$ & $open \quad object;$ \\
\multicolumn{2}{@{}l}{$\{\;<$$Open$$>$.objId} & $\, =$ &$object$.objId \\
\multicolumn{2}{@{}l}{$\;\;\,object$.objTp} & $\, =$ & $<$$Open$$>$.objTp $\qquad \qquad \;\,\}$\\
\rowcolor[gray]{.9} $(iv)$ & $<$$Read$$>$ & $::=$ & $receive \; object1 \leftarrow object2;$ \\
\multicolumn{2}{@{}l}{$\{\;<$$Read$$>$.varId} & $\, =$ &$object1$.objId \\
\multicolumn{2}{@{}l}{$\;\;\,object1$.objTp} & $\, =$ & $var$ \\
\multicolumn{2}{@{}l}{$\;\;\,object2$.objId} & $\, =$ & $<$$Read$$>$.objId \\
\multicolumn{2}{@{}l}{$\;\;\,object2$.objTp} & $\, =$ & $<$$Read$$>$.objTp $\qquad \qquad \;\;\}$\\
\rowcolor[gray]{.9} $(v)$ & $<$$Write$$>$ & $::=$ & $send \; object1 \rightarrow object2;$ \\
\multicolumn{2}{@{}l}{$\{\;<$$Write$$>$.varId} & $\, =$ &$object1$.objId \\
\multicolumn{2}{@{}l}{$\;\;\,object1$.objTp} & $\, =$ & $var$ \\
\multicolumn{2}{@{}l}{$\;\;\,object2$.objId} & $\, =$ & $<$$Write$$>$.objId \\
\multicolumn{2}{@{}l}{$\;\;\,object2$.objTp} & $\, =$ & $<$$Write$$>$.objTp $\qquad \qquad \}$\\
\rowcolor[gray]{.9} $(vi)$ & $<$$IntrlvRW$$>$ & $::=$ & $while(receive \; object1 \leftarrow object2;)\{$ \\
\rowcolor[gray]{.9}  & & & $\quad send \; object3 \rightarrow object4;$ \\
\rowcolor[gray]{.9}  & & & $\}$ \\ 
\multicolumn{2}{@{}l}{$\{\;object3$.objId} & $\, =$ & $object1$.objId \\
\end{tabular}
\noindent \begin{tabular}{@{}l@{$\;$ }l@{ }l@{ }l}
\multicolumn{2}{@{}l}{$\;\;\,object1$.objTp} & $\, =$ & $var$ \\
\multicolumn{2}{@{}l}{$\;\;\,object3$.objTp} & $\, =$ & $var$  \\
\multicolumn{2}{@{}l}{$\;\;\,object2$.objId} & $\, =$ & $<$$IntrlvRW$$>$.obj1Id \\
\multicolumn{2}{@{}l}{$\;\;\,object2$.objTp} & $\, =$ & $<$$IntrlvRW$$>$.obj1Tp \\
\multicolumn{2}{@{}l}{$\;\;\,object4$.objId} & $\, =$ & $<$$IntrlvRW$$>$.obj2Id \\
\multicolumn{2}{@{}l}{$\;\;\,object4$.objTp} & $\, =$ & $<$$IntrlvRW$$>$.obj2Tp $\qquad \quad \}$\\
\rowcolor[gray]{.9} $(vii)$ & $<$$DirectCopy$$>$ & $::=$ & $send \; object1 \rightarrow object2;$ \\
\multicolumn{2}{@{}l}{$\{\;<$$DrctCopy$$>$.obj1Id} & $\, =$ &$object1$.objId \\
\multicolumn{2}{@{}l}{$\;\;\,object1$.objTp} & $\, =$ & $<$$DrctCopy$$>$.obj1Tp \\
\multicolumn{2}{@{}l}{$\;\;\,<$$DrctCopy$$>$.obj2Id} & $\, =$ &$object2$.objId \\
\multicolumn{2}{@{}l}{$\;\;\,object2$.objTp} & $\, =$ & $<$$DrctCopy$$>$.obj2Tp $\qquad \quad\}$ \\
 \hline
\end{tabular}
\end{scriptsize}\\\\

\subsubsection{Propagation} Propagation differs from duplication by copying the data from the self-reference towards a communicating object. Consequently, propagation shows some syntactic similarities with duplication with additional readjustments to insert a potential format process: their main differences thus lie in the semantic rules. Two major modifications are brought to the start propagation production rule. The first modification replaces the permanent type for the target object by the communicating type ($<$$Propagate$$>$.targTp$\,=\,obj\_com$). A communicating object can be a network connection, a file shared over P2P folders or networks drives, or a simple mail. This example illustrates the importance of typing. The second modification specifies that the source of propagation can be either the auto-reference or the intermediate result of the duplication ($<$$Propagate$$>$.srcTp$\,= \,this \, \vee \, <$$Propagate$$>$.srcId $\,= \,<$$Duplicate$$>$.targId). These alternatives explain the disjunction of semantic equations.\\

\begin{scriptsize}
\noindent 
\begin{tabular}{@{}l@{$\;$ }l@{ }l@{ }l}
\hline
\hline
\rowcolor[gray]{.9} $(i)$ & $<$$Propagate$$>$ & $::=$ & $<$$Open$$><$$Read$$><$$Transmit$$>$\\
\rowcolor[gray]{.9} & & $\; \mid $ & $<$$Read$$><$$Open$$><$$Transmit$$>$\\
$\{$ & ... &  &  \\
\multicolumn{2}{@{}l}{$\;\;\,$($<$$Propagate$$>$.srcTp} & $\,=$ & $this$ \\
\multicolumn{2}{@{}l}{$\;\;\,\;\vee <$$Propagate$$>$.srcId} & $\,=$ & $<$$Duplication$$>$.targId) \\
 & ... &  &  \\
\multicolumn{2}{@{}l}{$\;\;\,<$$Propagate$$>$.targTp} & $\, =$ & $obj\_com$ \\ 
 & ... &  &  $\qquad \qquad \qquad \qquad \qquad \qquad\}$\\
\rowcolor[gray]{.9} $(ii)$ & $<$$Transmit$$>$ & $::=$ & $<$$Format$$><$$Write$$>$ \\
\rowcolor[gray]{.9} & & $\; \mid $ & $<$$Write$$>$\\
 \hline
\end{tabular}
\end{scriptsize}\\

\section{Translation into the language using abstraction}
	In the context of detection, a set of data giving information about the structure or the actions of the malicious code is statically or dynamically collected. According to the level of the collection mechanism, the completeness of the available data is strongly impacted and the nature of this data may vary from simple instructions to system calls along with their parameters. These collected data remain specific to a given platform and to the language in which a piece of malware has been coded (native, interpreted or macro code). A first translation layer is thus required to abstract the collected data into a generic behavioral language (see Section II). Translation of basic instructions, either arithmetic (move, addition, subtraction...) or control related (conditional, jump...), into operations of the behavioral language is an obvious mapping which does not require further explanation. With regards to API calls and their parameters, their translation into interactions and objects from the behavioral language turns out to be more complex. Consequently, the two following parts focus on these aspects.

	\subsection{API calls translation}

	For an arbitrary program to access any service or resource from its environment, the Application Programming Interfaces (API) constitute a mandatory point enforcing security and consistency between the different accesses \cite{OB}. In the common case of native code accessing kernel services from the operating system, API calls are also denoted system calls but the first notation will prevail to remain consistent. For each programming language the set of available API can be classified into distinct interaction operations. This set of API being finite and supposedly stable, the translation can be defined as a direct mapping over the space of the interaction classes, guaranteeing the completeness of the process. A part of this mapping is given in Table I for a subset of the Windows Native API (Ntdll: \cite{Nt}), as well as a subset of the VBScript API functions. The table has been refined according to the nature of the manipulated objects. The API name, on its own, is not always sufficient to determine the interaction class as the network example demonstrates.  Network managers and simple files use common API: for a clear distinction, the file path must also be interpreted (\texttt{$\backslash$device$\backslash$Afd$\backslash$Endpoint} under Windows). Sending and receiving operations then depends on the driver control code transmitted with \texttt{NtDeviceIoControlFile} (\texttt{IOCTL\_AFD\_RECV}, \texttt{IOCTL\_AFD\_SEND} \cite{Afd}). When required, parameters thus become additional inputs of the mapping: $\{API\;name\} \times (\{Parameters\}\cup \{\epsilon\}) \rightarrow \{Interaction \; class\}$.

\begin{table*}[!ht]
  \centering
  \includegraphics*[scale=0.58]{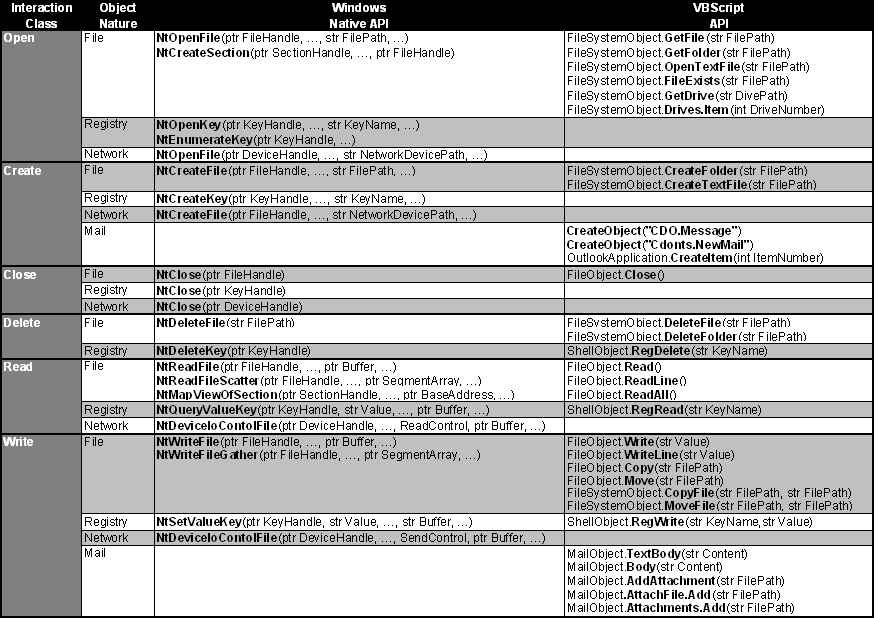} 
  \label{tab:apicalls}
%  \usefont{OT1}{cmr}{b}{n}
  \caption{Windows Native and VBScript API mapping to interaction classes.}
\end{table*}

	\subsection{Parameters interpretation}
	Parameters are important factors in interactions, not only to distinguish certain ambiguous classes of interactions like in the previous network example. Parameters also identify the different objects involved in interactions and assess their criticality through the typing of these objects. Interpretation of the parameters thus finishes the abstraction from the platform and language begun with the API translation. Due to their various natures, parameters can not be translated using a simple mapping like for API. Basically, decision trees are more adaptive tools capable of interpreting parameters according to their representation:
\begin{description}
\item[\textbf{Simple integers:}] $\qquad\qquad \;$ Integer attributes are mainly constants specific to an associated API. They mainly condition the interpretation of the interaction class of the API. For \texttt{NtDeviceIoControlFile}, the different IO control codes are typical examples. A simple hard-coded comparison is required to detect the main important constants.
\item[\textbf{Address and Handles:}] $\qquad\qquad \qquad\;\;$ Addresses and handles are mainly used to identify the different objects appearing in the collected data. These parameters are particularly useful to study the data flow between object. A variable for exemple may be represented by an address $a_v$ and a potential size $s_v$. Every address $a$ such as $a_v\leq a\leq a_v+s_v$ will refer to the  same variable. Certain addresses have important properties and may be refined by typing: import tables, services descriptor table, entry points... In order to interpret these specific elements, a possible decision tree based on the partitioning of the address space is proposed in the Figure 2 .
\item[\textbf{Character strings:}] $\qquad\qquad \quad\;$ String parameters contains the richer information about objects to interpret. Most of these parameters are paths satisfying a hierarchical structure where every element is important: from the root element identifying drives, drivers and registry paths, passing by the intermediate directories providing object localization, until the real name of the object. The hierarchical structure of the paths is well adapted for a progressive analysis which can once again be modelled as a decision tree. Such a tree is described in the Figure 3 for a more complex string interpretation.\\
\end{description}
	
\begin{figure}[!ht]
  \centering
  \includegraphics*[scale=0.4]{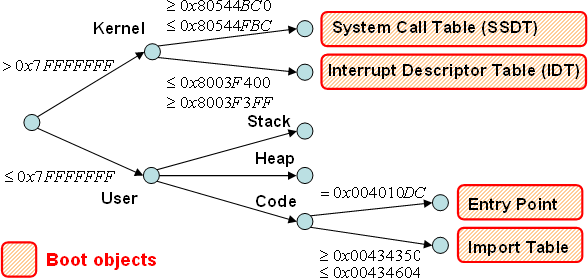} 
  \label{fig:address}
%  \usefont{OT1}{cmr}{b}{n}
  \caption{Addresses interpretation.}
\end{figure}

\begin{figure*}[!ht]
  \centering
  \includegraphics*[scale=0.45]{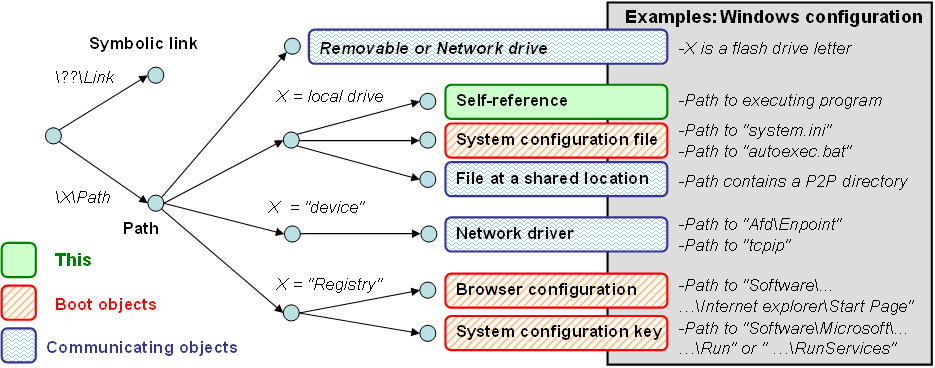} 
  \label{fig:string}
%  \usefont{OT1}{cmr}{b}{n}
  \caption{Character strings interpretation.}
\end{figure*}

	Building decision trees requires a precise identification of the critical resources of a system. A methodology, reproducible to various systems, is to proceed by considering successively the different system layers: hardware layer, operating system layer and applicative layer. For each layer, a scope must be defined encompassing the significant components; the resources involved either in the installation, the configuration or the use of these components must then be analyzed for potential misuse:
\begin{description}
\item[\textbf{Hardware layer:}] $\qquad\qquad \;$ For the hardware layer, the scope has been restricted to the different interfaces open to external locations (Network, CD, USB). With respect tothe conditions of use of these interfaces, the main resources to consider remain the drivers used to communicate with the interfaces. Additional configuration files must also be considered because they may impact the connection parameters ("host" file) or the booting of the external element ("autorun.inf" for example).
\item[\textbf{Operating system layer:}] $\qquad\qquad\qquad\quad\;\;$ The configuration of the operating system is critical but is unfortunately dispersed in various locations (files, structures and tables in memory, registry keys). The scope is proportionally broadened. However, the critical resources are already well identified considering important aspects such as the boot sequence or the intermediate structures used to access the services and resources provided by the operating system (file system, process table, system call table...).    
\item[\textbf{Applicative layer:}] $\qquad\qquad \quad$ It is obviously impossible to consider all existing applications. To restrict the scope, the analysis must consider only connected and widely deployed applications (web browsers, mail clients, peer-to-peer clients, messaging, IRC clients). This restriction makes sense since malware, with regards to their propagation and interoperability requirements, must operate on a scale of platforms as large as possible. Once again for these applications, the resources involved in the communication (connections, transit locations) as well as in the configuration (application launch), must be considered.
\end{description}

	This identification of the critical resources used by malware is a complex manual configuration step, but nevertheless necessary. This identification process is however less cumbersome than analyzing the thousands of malware samples collected every day. Once these critical resources pinpointed, the configuration can be retrieved in a partially automated way. For example, the nature of the different drives can be recovered automatically (local, network, removable media). Similarly, the different installed peer-to-peer clients can be detected and their shared directories recovered. The localization of the system call and interrupt tables can also be recovered and so on. From our opinion, a full automation of the parameter interpretation may be very hard to achieve. With regards to intrusion detection, some attempts at fully-automated analysis of the system call parameters have been put forward for anomaly-based detectors \cite{KMVV03}. Their parameter interpretation relied on some deviation measurements from a legitimate model based on string length, character distribution and structural inference. These factors are significant for intrusion mainly because most attacks used misformatted parameters to take advantage from a weakness of the system implementation. This approach should prove less efficient with malware since they mainly use legitimate or legitimate appearing parameters. Moreover these anomaly-based approaches do not explain the use of the object for the malware; this would require an additional manual analysis. Fortunately, critical resources for a given platform are quite restricted and can potentially be enumerated. Thus, parameter interpretation by decision trees seems a good trade off between partial automation and manual analysis of the platform facilities.

\section{Detection using parsing automata}
	As a consequence of the attribute-grammar formalization, detecting malicious behaviors can be reduced to the problem of parsing their grammatical description. Before going any further, we must state two important hypothesises. These grammars must be both LL grammars and L-attribute grammars: attribute dependencies are only allowed from left to right in the production rules. These hypothesises are important since they condition the fact that syntactic parsing and attribute evaluation can be achieved in a single pass. These properties are not necessarily satisfied by the MBL generative grammar but they prove true for the sub-grammars describing the different malicious behaviors (see Section II). Detection can then be implemented by LL-parsers responsible for the top-down construction of annoted leftmost-derivation trees. LL-parsers are basically pushdown automata with attribute evaluation as given in the Definition 2 below. \\
	
\begin{defi} A LL-parser is a particular pushdown automaton $A$ that can be built as a ten-tuple\\
 $<$$Q,\Sigma , D, \Gamma_p, \Gamma_s,\delta ,q_0,Z_{p,0},Z_{s,0},F$$>$  where:\\
- Q is the finite set of states,\\
- $\Sigma $ is the alphabet of input symbols and $D$ the set of possible values for attributes,\\
- $\Gamma_p$ is the alphabet of the parsing stack and $\Gamma_s$ the alphabet of the semantic stack,\\
- $\delta $ is the transition function of the form $Q\times (\{\Sigma\cup \epsilon\},D^*) \times (\Gamma_p,\Gamma_s)  \rightarrow Q\times (\{\Gamma_p \cup \epsilon\},\Gamma_s)$ which defines both the production rules and the semantic routines,\\
- $q_0 \in Q$ is the initial state,\\
- $Z_{p,0}$ and $Z_{s,0}$ are respectively the initial symbol of the parsing and semantic stacks,\\
- $F\subset Q$ is the set of accepting states.\\
\end{defi}
		
	Because several behaviors are monitored, a dedicated automaton is deployed in parallel for each of these behaviors. A single automaton can parse several instances of the behavior storing its progress in independent states and stacks. For each read input, all automata progress along their derivation trees at the same time as described in the Figure 4. When an irrelevant input is read (an operation interleaved inside the behavior for example), instead of causing an error state in the automata, this input is simply dropped. The global algorithm is procedurally defined below:\\[0.5em]
\begin{small}
\textbf{Procedure:} BehaviorDetection($e_1$,...,$e_t$)\\
where $e_i$ are collected events (input symbol combined with semantic values: $(\{\Sigma\cup \epsilon\},D^*)$) \\
\indent \textbf{For each} collected event $e_i$ \textbf{do:}\\
\indent $\;\mid$ \indent \textbf{For all} the automata $A_k$ \textbf{from} $k=1$ \textbf{to} $k\leq n$ \textbf{do:} \\
\indent $\;\mid$ \indent $\;\mid$ \indent m = number of derivation; \\
\indent $\;\mid$ \indent $\;\mid$ \indent \textbf{For all} state and stack triple $(Q_{k,j},\Gamma_{pk,j},\Gamma_{sk,j})$\\
\indent $\;\mid$ \indent $\;\mid$ \indent \textbf{from} $j=1$ \textbf{to} $j\leq m$ \textbf{do:} \\
\indent $\;\mid$ \indent $\;\mid$ \indent $\;\mid$ \indent $A_k$.\textbf{ll-parse}($e_i$,$(Q_{k,j},\Gamma_{pk,j},\Gamma_{sk,j})$)\\
\indent $\;\mid$ \indent $\;\mid$  \indent \textbf{Next}\\
\indent $\;\mid$  \indent \textbf{Next}\\
\indent \textbf{Next}\\[0.5em]
\textbf{Procedure:} A.ll-parse($e$,$(Q,\Gamma_{p},\Gamma_{s})$)\\
\indent \textbf{If} $e$, $Q$, $\Gamma_{p}$, $\Gamma_{s}$ match a transition $T\in \delta_A$ \textbf{Then} \\
\indent  $\;\mid$ \indent \textbf{If} $e_i$ introduces a possible ambiguity\\
\indent  $\;\mid$ \indent  \textbf{Then} duplicate state and stack triple $(Q,\Gamma_{p},\Gamma_{s})$\\
\indent  $\;\mid$ \indent Compute transition $T$ to update $(Q,\Gamma_{p},\Gamma_{s})$\\
\indent  $\;\mid$ \indent \textbf{If} $Q$ is an accepting state $Q\in F_A$\\
\indent  $\;\mid$ \indent \textbf{Then} Malicious behavior detected\\
\indent  \textbf{Else} ignore $e_i$\\
\indent  \textbf{End if}\\[0.5em]
\end{small}
\textbf{Listing:} Detection algorithm based on parallel automata.\\

\begin{figure}[!ht]
  \centering
  \includegraphics*[scale=0.38]{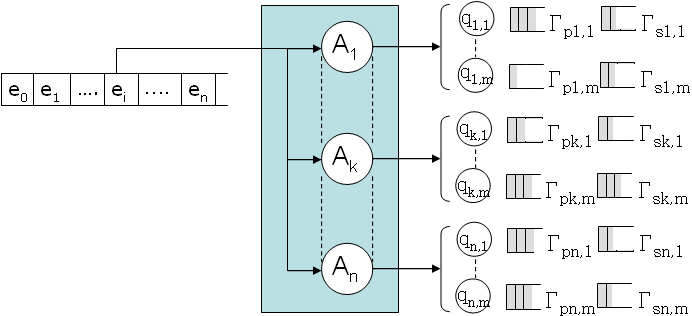} 
  \label{fig:parallelautomata}
  %\usefont{OT1}{cmr}{b}{n}
  \caption{Parallel automata for behavioral detection. %\usefont{OT1}{cmr}{m}{n} 
The collected events $e_i$ containing the input symbols and semantic values are fed into the parsing parallel automata $A_k$. Each of these automata manages a set of independent states $q_k$, parsing and semantic stacks $\Gamma_{pk}$, $\Gamma_{sk}$ corresponding to different derivations.}
\end{figure}

\subsection{Semantic routines for prerequisites and consequences}
	A malicious behavior is basically a sequence of operations where each operation prepares for the next one. Looking at recent works in intrusion detection, an intrusion scenario has been defined similarly as a sequence of dependent attacks \cite{CM02,NCRX04}. For each attack to occur, a set of prerequisites or preconditions must be satisfied. When completed, the attack generates new consequences, also called postconditions. In a formalization of malicious behaviors by attribute grammars, the sequence order is driven by the syntax whereas prerequisites and consequences can be directly modelled by semantic rules of the form $Y_i.\alpha = f(Y_1.\alpha_1...Y_n.\alpha_n)$ (see Definition 1).
\begin{description}
\item[\textbf{Checking prerequisites:}]  $\qquad\qquad\qquad\qquad$ Prerequisites are defined by specific semantic rules where the attributes at the left side of the equation are attached to a terminal symbol ($Y_i \in \Sigma$). During parsing a semantic value is collected with the input symbol and transmitted to the automaton. This value is then compared to the computed value using the inherited and already synthesized attributes. The comparison correspond to the matching step on the semantic stack $\Gamma_s$ performed during a transition from $\delta$.
\item[\textbf{Evaluating consequences:}] $\qquad\qquad\qquad\qquad\;\,$ When the left-side attribute is attached to a non-terminal ($Y_i \in V$), and all right-side inherited and synthesized attributes are valued, the new attribute can be evaluated. The evaluation corresponds to the reduction step performed during a transition from $\delta$: this transition results in a modification of the semantic stack $\Gamma_s$ where the new value has been pushed.
\end{description}

\subsection{Ambiguity support}
	In the detection algorithm, all events are confronted to the behavior automata. However these events may be sometimes unrelated to the behavior or unuseful to complete the behavior. Unrelated events do not match any transition in the parsing procedure and are simply dropped. But this measure reveals itself insufficient for unuseful events which raise ambiguities in the sense that they may be related to the behavior but parsing them makes the derivation fail unpredictably. Let us take an explicit example in the case of duplication: after opening the self-reference, two files are consecutively created. If duplication is achieved between the self-reference and the first file, parsing will succeed. On the opposite, if duplication is achieved between the self-reference and the second one, parsing will fail. This can be explained by the fact that the automaton will progress with the first creation beyond the state of accepting a second creation. The algorithm should thus be able to manage the different file combinations. Similar ambiguities may be observed with the data-flows between variables and objects. \\
\indent Ambiguities are handled by the detection algorithm using duplication of the derivation. Before the reduction of a transition, if the operation is potentially ambiguous, the current state of the derivation is stored in a new triple containing the current state and the parsing and semantic stacks. This way, the algorithm can handle the different combinations of events without a complete mechanism of backtracking to come back and forth in the derivation trees, which would have proved too cumbersome for detection in real-time.

\subsection{Time and space complexity}
	The whole process of LL-parsing is known to be linear in time in function of the number of parsed symbols. Because of ambiguities, the detection algorithm has a greater complexity. Let us consider that a single call to the parsing procedure is the reference operation. This procedure is decomposed in three basic blocks: matching, reduction and accept, which can be resumed to two comparisons and a computation. Let us reason in the worst case scenario, that is to say that, all collected events are related to the different behavior automata and all these related events potentially introduce ambiguities. On the opposite, in the best case scenario, no ambiguity is raised. The resulting complexities are given in the Proposition 1 below.\\
	
\begin{prop} In a worse case scenario, behavioral detection using attributed automata has a time complexity in $\vartheta(k(2^n-1))$ and a space complexity in $\vartheta(k2^{n}(2s)$ where $k$ is the number of automata, $n$ is the number of input symbol and $s$ is the maximum stack size. In a best case scenario, time complexity drops to linear time $\vartheta(kn)$ and space complexity becomes independent form the number of inputs $\vartheta(k2s)$.\\
\end{prop} 

	The worst case complexity may seem important but it quickly drops when the number of ambiguous events decreases. The experimentations presented in the section V. will show that the ratio of ambiguous events is limited and the algorithm offers satisfactory performances. These experimentations will also show that an important ratio of ambiguous events is already a sign of malicious activity. Based on these results, a new assessment of the average practical complexity will be provided.\\
	
\begin{proof}
	In a best case scenario, the number of derivation for each automaton is constant and remains equal to one. Considering the worse case scenario, all events are potentially ambiguous for all the current derivations. Technically, this ambiguity multiplies by two the number of derivations at each iteration of the main loop. Consequently, each automaton realises $2^{i-1}$ times the loop on the different derivations at the $i^{th}$ iteration of the main loop. The time complexity is then equivalent to the number of calls to the parsing procedure:\\[0.3em]
\begin{tabular}{lll}%{l@{$\;$ }l@{ }l@{ }ll}
$(1) \; k+2k+...+2^{n-1}k$ & $=$ & $k(1+2+...+2^{n-1})$ \\
 & $=$ & $k(2^n - 1)$ \\
\multicolumn{3}{l}{(Power sumation of the $n$ first powers of 2)}\\[0.3em]
\end{tabular}

The maximum number of derivations is reached after the last iteration where all automata manage $2^{n}$ parallel derivations. Each derivation is stored in two stacks of size $s$. This moment coincide with the maximum occupation of the memory space:\\[0.3em]
\begin{tabular}{llll}%{l@{$\;$ }l@{ }l@{ }ll}
$(2) \; k2^{n}(2s)$.\\
\end{tabular}
\end{proof}

\section{Prototype implementation}
As presented in the previous section, we have developed a prototype in two layers: a specific layer combining data collection and abstraction and a second generic layer for detection. The overall architecture is described in the Figure 5. As a proof of concept, the abstraction layer has been implemented for two different languages: native code from PE executables and interpreted Visual Basic Script. The detection layer is then generically deployed above without any adaptation to the origin of the data. The different elements of the architecture are described in the next sub-sections.
\begin{figure}[!ht]
  \centering
  \includegraphics*[scale=0.4]{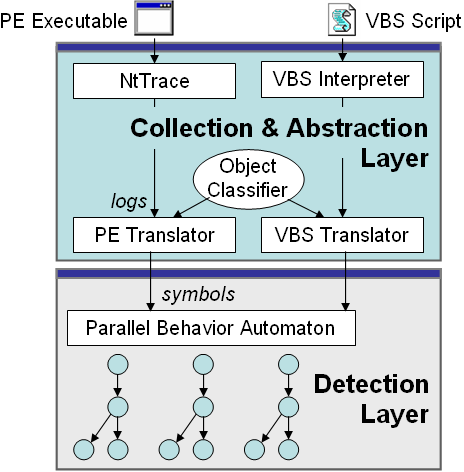} 
  \label{fig:architecture}
%  \usefont{OT1}{cmr}{b}{n}
  \caption{Detector architecture combining detection and abstraction layers. The abstraction layer is directly linked to the collection mechanism. Each component of the layer interprets the specific features of the languages whereas the object classifier is common to all and interprets the specificity of the platform. Above this layer, a second detection layer based on parallel behavioral automata parses the interpreted data independently from their original source. }
\end{figure}

\subsection{Analyzer of process traces}
	Process traces provide useful information about the system activity of an executable. Several tools exist to collect such traces. Collection mechanisms are out of the scope of this article and no collection tool has been reimplemeted. The prototype simply uses an existing tool to collect Windows native calls, their arguments and returned  values called NtTrace \cite{Nttrace}.\\
\indent Contrary to static analysis, the main point with dynamic collection mechanisms (real-time or emulation based) is that most behaviors are conditioned by external objects and events: available target for infection or listening servers for network propagation for example. In order to increase the mechanism coverage and collect conditioned behaviors, a virtual environment has been used satisfying most of the triggering conditions as pictured in Figure 6. The trace collection has been deployed over Qemu \cite{Qemu} using an image of a Windows XP installation where useful services and resources have been configured: ISP account, mail client, P2P client, different potential targets (executable files, pictures, music, html pages). Outside the virtual machine, different server emulations have been deployed (DNS, SMTP). These servers are not directly used to collect data but their presence is mandatory to establish a connexion and capture any network activity at the system call level. A last remark is that malware can become inactive beyond a predefined date, the system time has thus been reconfigured in a previous date.\\
\indent Translation is then deployed line by line on the collected traces. It directly implements the results from Section III for API call translation and parameter interpretation. Referenced API are directly classified according to Table 1 and an object classifier, specifically designed for a Windows configuration, is then called on the parameters. The API unreferenced in Table 1 are simply ignored for the moment, until their integration in a future version. As for the object classifier, it directly embeds decision trees such as the ones described in the Figures 2 and 3, but with more complete information.\\
\indent Looking specifically at creation and opening interactions, when resolved, a correspondence is established between the name of the involved objects and their references (addresses, handles); the result is then stored in a database. Following interactions check for these references during interpretation. Conversely, on deleting and closing interactions, this correspondence is destroyed for the remainder of the analysis. Names and identifiers must be unlinked since a same address or handle number could be reused for a different object.\\
\indent During this translation process, sequences of identical operations as well as sequences of two combined operations are detected and formatted into a loop in order to decrease the log size. 
	
\begin{figure}[!ht]
  \centering
  \includegraphics*[scale=0.36]{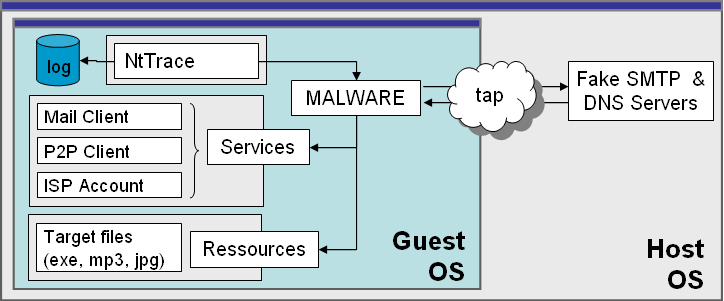} 
  \label{fig:pecollector}
%  \usefont{OT1}{cmr}{b}{n}
  \caption{Malware API calls collection environment. }
\end{figure}

\subsection{Analyzer of Visual Basic Scripts}
	With regards to Visual Basic Script (VBScript), no collection tool such as NtTrace was available. Unfortunately, VBScript is proprietary explaining that the few existing parsers and interpreters remain commercial. We have thus developed our own collection mechanism to directly embed the abstraction layer inside, which was not feasible in a commercial product. In addition, by developing only a partial interpreter with restricted code execution, we have been able to increase the performance of the instrumentation.\\
\indent VBScript being an interpreted language (high-level language), its analysis is simpler than native code because of the visibility of the source code but also because of some integrated safety properties: no direct code rewriting during execution and no arbitrary transfer of the control flow \cite{MR08}. For these reasons, path exploration becomes conceivable. To do so, we have divided the analyzer in two parts: a first static part collecting different information on the script structure and normalizing the code to fight obfuscation, and a second dynamic part exploring the different execution paths and collecting significant events. In addition, the object classifier has been integrated in order to type the event-related objects.\\

\subsubsection{Static syntactic analyzer} Syntactic analysis heavily depends on the specification of the VBScript language \cite{VBS08}. Before unfolding the different syntactic rules, the script file is quickly parsed to localize and retrieve the signatures of the local functions and procedures. The script file is then parsed line by line to collect important information which are stored in a global structure described in the Figure 7. The declared variables and constants are recovered from the \texttt{"Dim"} and \texttt{"Const"} derivations. The analyzer also identifies the important managers declared in the script: file managers (\texttt{"Scripting.FileSystemObject"}), shell managers (\texttt{"WScript.Shell"}), network managers (\texttt{"WScript.Network"}) or mail managers (\texttt{"Outlook.Application"},\texttt{"CDO"}).\\
\indent These declaration lines as well as comment lines are tagged to avoid any double analysis; other lines are normalized and stored in the script structure. The first step of code normalization is to remove the numerous syntactic shortcuts provided by VBScript: for example, the single-line concatenated instructions using \texttt{":"} are dispatched on independent lines, or the \texttt{"With"} structure applied to a given object is reversed by concatenating this object in head of the lines starting with a method access. Normalization is also critical to thwart obfuscation. Current obfuscation techniques consist in splitting the different strings in several substrings; characters may then be encoded into integers using the \texttt{"Chr"} primitive. Normalization reverses the process by decoding the integers into characters and concatenating consecutive substrings into a single one. The whole process is described in the Figure 8. Obfuscation is also achieved in some scripts by string encryption. String encryption techniques in VBScript remain quite basic since the algorithm must work from the set of printable characters to the exact same set for the ciphered text. This explains that most algorithms are simply based on permutations. During the static analysis, the decryption routine is localised and copied in a script file. This routine is finally called on-demand with the right parameters to decipher the different encrypted strings. \\

\begin{figure}[!ht]
  \centering
  \includegraphics*[scale=0.3]{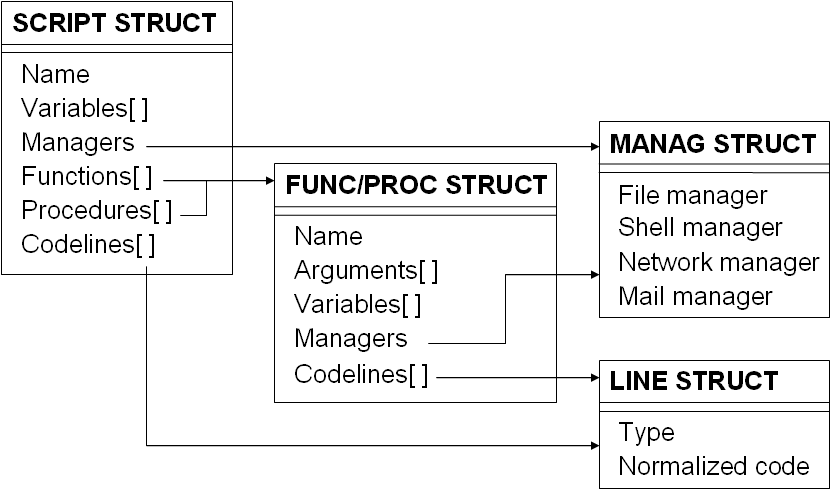} 
  \label{fig:scriptstructure}
%  \usefont{OT1}{cmr}{b}{n}
  \caption{Structure of a VBScript file. This global structure stores the important informations concerning the script: variables, managers, functions and procedures. It also stores the normalized code for execution exploration.}
\end{figure}

\begin{figure}[!ht]
  \usefont{OT1}{cmtt}{m}{n}
  \small{
  $(i)$ execute "set QAI5NPN1 =T228IV93." \& Chr(65) \& Chr(116) \& Chr(116) \& Chr(97) \& Chr(99) \& Chr(104) \& Chr(109) \& Chr(101) \& Chr(110) \& Chr(116) \& Chr(115)\\
  $(ii)$ execute "set QAI5NPN1 =T228IV93." \& "A" \& "t" \& "t" \& "a" \& "c" \& "h" \& "m" \& "e" \& "n" \& "t" \& "s"\\ 
  $(iii)$ execute "set QAI5NPN1 =T228IV93.Attachments"
  }
  \label{fig:reverseobf}
%  \usefont{OT1}{cmr}{b}{n}
  \caption{Reversing obfuscation by code normalization. This portion of obfuscated code has been extracted from a VBSWG worm variant. Code normalization is achieved by several steps: encoded integers are restored as characters $(i-ii)$, splitted strings are then concatenated $(ii-iii)$.}
\end{figure}

\subsubsection{Dynamic interpreter:} A partial script interpreter has been defined to explore the different execution paths. This interpreter is only partial in the sense that the script code is not really executed but only significant operations and dependencies are collected. The code is processed line by line with detection of the numerous syntaxes for conditional (\texttt{"If...Then...Else"}, \texttt{"Switch"}) and loop (\texttt{"While"}, \texttt{"For"}, \texttt{"For each"}) structures in order to explore the different possible paths. Calls to local procedures and functions are also addressed by saving the current position and jumping inside their code, however, with a particular restriction: recursive calls have been blocked to avoid any stack overflow. Both recursive calls and mutual recursive calls involving multiple functions and procedures are detected by managing a call stack.\\

\begin{figure*}[!ht]
  \centering
  \includegraphics*[scale=0.35]{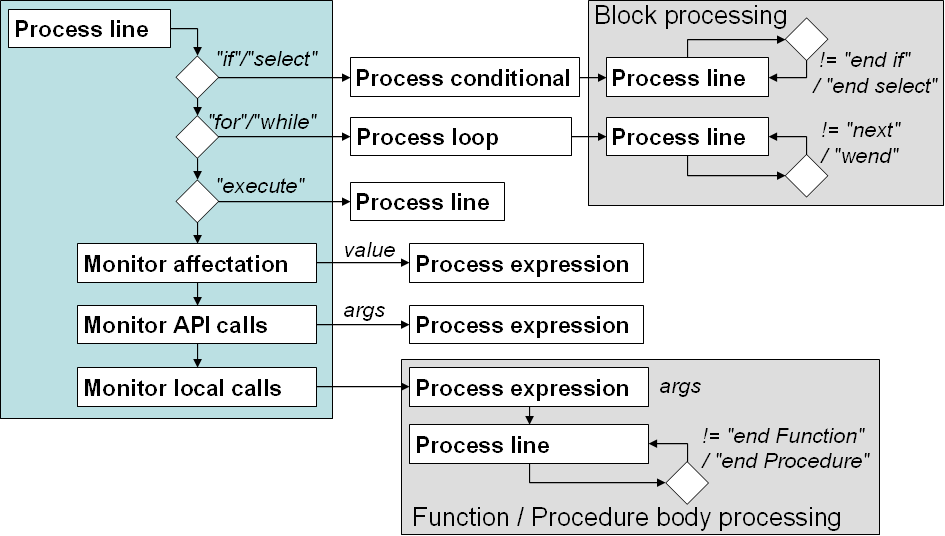} 
  \label{fig:lineprocess}
%  \usefont{OT1}{cmr}{b}{n}
  \caption{Articulation of the different levels of processing. This articulation is directly linked to the script structure in code blocks chained according to the control flow. }
\end{figure*}

\indent Each script line is processed to retrieve the different monitored API calls manipulating files, registry keys, network connections or mails. The monitored calls are classified according to Table 1 from Section III. In addition, variable affectations are also important for the data-flow and are thereby monitored. All these operations require a second level of analysis to process the expressions used as arguments or values for affectations. The global articulation between the different levels of processing is schematically described in the Figure 9.\\
\indent With regards to expression processing, in case of a single element, the resolution is immediate. However, it becomes more complicated with concatenated values (\texttt{"\&"}): the different elements of the expressions are analyzed and only the element with the greater type is kept as reference (see the Figure 1 for the type poset). This selection is used to decrease the number of data to monitor while focusing on more significant elements. Imbricated calls may also be encountered inside expressions under two forms: either $res = call_1(arg_1).call_2(arg_2)$ or $res = call_1(call_2(arg_2),arg_1)$. In these particular cases, a new intermediate object is created to store the result of the call. Using this newly created object, a new line is then built before to be processed like any other: respectively $int = call_1(arg_1) \;/\; res = int.call_2(arg_2)$ and $int = call_2(arg_2) \;/\; res = call_1(int,arg_1)$. The intermediate object preserves the data-flow during the analysis. Generally speaking the data-flow is really important and the different references and aliases for objects must be followed up through the processing of expressions, and in particular at some key operations:
\begin{description}
\item[\textbf{Local function/procedure call:}]  $\qquad\qquad\qquad\qquad\qquad\;$ Before jumping inside a function or a procedure, the referring names of the arguments must be added as references for the objects passed as parameters. These names are actually recovered from the static analysis of the signature. Once the whole code executed, the added references must be removed to prepare for a next call. In addition, in the case of functions, the returned value must be associated to the result variable. In VBS, the return value is stored under an object named liked the function. Once this value stored, the function name must be removed from the object reference.
\item[\textbf{Monitored API call:}]  $\qquad\qquad\qquad\;$ The API is first classified according to the operation classes but the API name also indicates the natures of the involved objects. After an API call, the references are updated for these objects as well as their type. In case of a new object, it is typed for the first time using the object classifier, otherwise; its type is refined according to its newly discovered nature.
\item[\textbf{Affectation:}]  $\qquad\quad$ When an affectation occurs, the affected value is first processed as an expression and the references of the affected object must be updated with the result. String processing is also very common in VBS (\texttt{"Mid, Left, Right, LTrim, RTrim, UCase, LCase, Replace..."}): it has been treated as a special case of affectation to avoid any loss of the data flow.
\item[\textbf{Call to execute:}]  $\qquad\qquad\;$  The following expression must be evaluated before to be written down in a newly created line for processing.\\
\end{description}

\subsubsection{Object classifier refinement}
The same object classifier than for NtTrace can be reused like pictured in the Figure 5. VBScript being mainly based on character strings, the address classifier part is actually unused. However, extensions to the string classifier have been brought to best fit the script particularities. In first place, constants linked to VBScript have been added like \texttt{"Wscript.ScriptName"} and \texttt{"ScriptFullName"} for the self-reference. In addition scripts are launched differently from executables offering new way to start automatically. For this reason, boot objects have been refined such as the \texttt{"Start page"} from \texttt{"Internet Explorer"} and configuration files such as \texttt{"script.ini"} for \texttt{"mIRC"} in case of IRC worms.\\
\indent An important precision must be brought with regards to classification: as already said, the nature of an object may affect typing. According to the poset from the Figure 1, a variable can not be typed as the self-reference. This consideration has helped to avoid a few false positives where a variable containing the name of the script was written down in a log file, which is common in scripts.

\subsection{Detection automata}
	The transitions corresponding to the different grammar production rules have directly been coded in a prototype. The real implementation is similar to the algorithm presented in Section III.\\
\indent Only a few enhancements have been brought to the algorithm in order to increase the performance. The first enhancement is a mechanism to avoid duplicate derivations. Coexisting derivations with identical states and stacks would only artificially increase the number of algorithm iterations without identifying other behaviors than the ones already detected. The second enhancement is related to the close and delete operations on objects. Once again, in order to decrease the numer of iterations, the derivations where no interactions intervene between the opening/creation and the closing/deletion of a same related object are destroyed. These two mechanisms have proved helpful in regulating the number of parallel derivations.

\section{Experimentation and discussions}
	For experimentation, we have gathered hundreds of samples to confront to our prototype. The pool of samples is divided into two categories: the Portable Executables and the Visual Basic Scripts. For each category, about 200 malware and 50 legitimate samples were gathered. We have considered different types of malware and legitimate applications as descibed by the pool repartition in the Figure 11. These malware were mainly downloaded from two main sources: \cite{VX} and \cite{OC}, whereas legitimate samples have been selected from an healthy system installation.\\% (Pour NtTrace tester le moteur polymorphique fonctionnel).\\
	
\begin{figure*}[!ht]
  \centering
  \includegraphics*[scale=0.5]{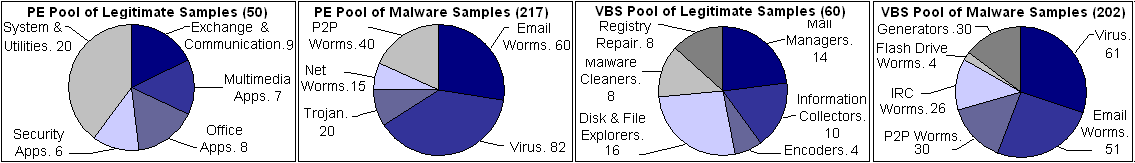} 
  \label{fig:samples}
%  \usefont{OT1}{cmr}{b}{n}
  \caption{Repartition of the test pool. }
\end{figure*}

\subsection{Coverage}
	The experimentation has provided significant results with a detection rate of 51\% for PE executables and up to 89\% for VB Scripts. The detection rates, behavior by behavior, are wholly described in the tables II and IV. According to these tables, duplication is indeed the most significant malicious behavior. However the additional behaviors, and in particular residency, have helped to detect additional malware where duplication was missed. On the opposite, the false positive rates presented in the table III and V are almost inexistent. The only observed false positive can be easily explained: the given script is a malware cleaner which reinitializes the Internet Explorer start page after the infection. \\
\indent By closely observing the results behavior by behavior, some important false negative spikes can be localized in the PE results (Table II): duplication detection rate for PE Virii and propagation detection rate for Net and Mail Worms are typical examples. In fact, these phenomenons can mainly be explained by existing limitations in the collection mechanisms. The impact of the collection mechanism on the detection is not described here but is assessed in the dedicated next section VI.B.\\
\indent By comparison between the VB scripts and the PE traces, the obtained rates of false negatives are lower for VB scripts. This can also be explained by the coverage of the collection mechanism: the VB script analyzer works statically and has been developed on purpose; it is thus more complete. The remaining false negatives can be explained by two problems. The first is the ciphering of the whole malware body which is not supported yet. This problem can be corrected in later versions of the analyzer. The second problem is the cohabitation in a same page of JavaScript and VB script code which makes the syntactic analysis fail. An additional code localisation mechanism can be added to circumvent the problem.\\
%Morales: Mail:60,56%   P2P:66,20%  Virus:18,30%  Net:63,38%
\indent Globally, with regards to existing works analyzing system calls \cite{MCD08}, the observed detection rates for duplication are consistent with the results previously obtained. In fact, the real enhancements from this work are two folds. The first enhancement is the parallel detection of additional behaviors described in the same language: propagation, residency and overinfection. The second enhancement is the possibility to feed detection with data from other sources such as those coming from the script analyzer.\\
	
\begin{table*}
\centering
\begin{footnotesize}
\begin{tabular}{|l@{ }|l@{ }|l@{ }|l@{ }|l@{ }|l@{ }|r@{ }|}
\hlinewd{1.2pt} 
\rowcolor[gray]{0}\textcolor{white}{Behaviors} & \textcolor{white}{EmW} 
& \textcolor{white}{P2PW} & \textcolor{white}{V} & \textcolor{white}{NtW} & \textcolor{white}{Trj} & \textcolor{white}{Global}\\
\hlinewd{1.2pt}
\rowcolor[gray]{.8}[5.5pt][2.5pt] Duplication & 41(68,33\%) 
& 31(77,5\%) & 15(18,29\%) & 8(53,33\%) & 6(30\%) & 46,54\% \\ 
\hlinewd{1.2pt}
 By direct copy & 0(0\%) 
& 0(0\%) & 0(0\%) & 0(0\%) & 0(0\%) & 0,00\% \\
 By single read/write & 41(68,33\%) 
& 30(75\%) & 14(17,07\%) & 8(53,33\%) & 6(30\%) & 45,63\%\\
 By interleaved read/write & 9(15\%) 
& 3(7,5\%) & 3(3,66\%) & 3(0,2\%) & 0(0\%) & 8,29\%\\
\hlinewd{1.2pt}
\rowcolor[gray]{.8}[5.5pt][2.5pt] Propagation & 4(6,67\%) 
& 19(47,5\%) & 3(3,66\%) & 1(6,67\%) & 0(0\%) & 12,44\% \\ 
\hlinewd{1.2pt}
 By direct copy & 0(0\%) 
& 0(0\%) & 0(0\%) & 0(0\%) & 0(0\%) & 0,00\% \\
 By single read/write & 4(6,67\%) 
& 19(47,5\%) & 3(3,66\%) & 1(6,67\%) & 0(0\%) & 12,44\% \\ 
 By interleaved read/write & 0(0\%) 
& 0(0\%) & 0(0\%) & 0(0\%) & 0(0\%) & 0,00\% \\
\hlinewd{1.2pt}
\rowcolor[gray]{.8}[5.5pt][2.5pt] Residency & 36(60\%) 
& 22(55\%) & 5(60,98\%) & 6(40\%) & 9(45\%) & 35,94\%\\ 
\hlinewd{1.2pt}
\rowcolor[gray]{.8}[5.5pt][2.5pt] Overinfection test & 0(0\%) 
& 0(0\%) & 0(0\%) & 0(0\%) & 0(0\%) & 0,00\% \\
\hlinewd{1.2pt}
by conditional 1 & 0(0\%) 
& 0(0\%) & 0(0\%) & 0(0\%) & 0(0\%) & 0,00\% \\
by inverse conditional 2 & 0(0\%) 
& 0(0\%) & 0(0\%) & 0(0\%) & 0(0\%) & 0,00\% \\
\hlinewd{1.2pt}
\rowcolor[gray]{.8}[5.5pt][2.5pt] Global detection & 43(71,67\%) 
& 33(82,50\%) & 16(19,51\%) & 8(53,33\%) & 11(55,00\%) & 51,15\%\\ 
\hlinewd{1.2pt}
\end{tabular}
\end{footnotesize}
 \caption{PE Malware detection rates (EmW = Email Worms, P2PW = Peer-to-peer Worms, V = Virii, NtW = Net Worms, Trj = Trojans).}
\end{table*}

\begin{table*}
\centering
\begin{footnotesize}
\begin{tabular}{|l@{ }|l@{ }|l@{ }|l@{ }|l@{ }|l@{ }|r@{ }|}
\hlinewd{1.2pt} 
\rowcolor[gray]{0}\textcolor{white}{Behaviors} & \textcolor{white}{ComE} 
& \textcolor{white}{MM} & \textcolor{white}{Off} & \textcolor{white}{Sec} & \textcolor{white}{SysU} & \textcolor{white}{Global}\\
\hlinewd{1.2pt}
\rowcolor[gray]{.8}[5.5pt][2.5pt] Duplication& 0(0\%) 
& 0(0\%) & 0(0\%) & 0(0\%) & 0(0\%) & 0,00\% \\
\hlinewd{1.2pt}
 By direct copy& 0(0\%) 
& 0(0\%) & 0(0\%) & 0(0\%) & 0(0\%) & 0,00\% \\
 By single read/write& 0(0\%) 
& 0(0\%) & 0(0\%) & 0(0\%) & 0(0\%) & 0,00\% \\
 By interleaved read/write& 0(0\%) 
& 0(0\%) & 0(0\%) & 0(0\%) & 0(0\%) & 0,00\% \\
\hlinewd{1.2pt}
\rowcolor[gray]{.8}[5.5pt][2.5pt] Propagation& 0(0\%) 
& 0(0\%) & 0(0\%) & 0(0\%) & 0(0\%) & 0,00\% \\
\hlinewd{1.2pt}
 By direct copy & 0(0\%) 
& 0(0\%) & 0(0\%) & 0(0\%) & 0(0\%) & 0,00\% \\
 By single read/write& 0(0\%) 
& 0(0\%) & 0(0\%) & 0(0\%) & 0(0\%) & 0,00\% \\
 By interleaved read/write & 0(0\%) 
& 0(0\%) & 0(0\%) & 0(0\%) & 0(0\%) & 0,00\% \\
\hlinewd{1.2pt}
\rowcolor[gray]{.8}[5.5pt][2.5pt] Residency& 0(0\%) 
& 0(0\%) & 0(0\%) & 0(0\%) & 0(0\%) & 0,00\% \\
\hlinewd{1.2pt}
\rowcolor[gray]{.8}[5.5pt][2.5pt] Overinfection test & 0(0\%) 
& 0(0\%) & 0(0\%) & 0(0\%) & 0(0\%) & 0,00\% \\
\hlinewd{1.2pt}
by conditional 1 & 0(0\%) 
& 0(0\%) & 0(0\%) & 0(0\%) & 0(0\%) & 0,00\% \\
by inverse conditional 2 & 0(0\%) 
& 0(0\%) & 0(0\%) & 0(0\%) & 0(0\%) & 0,00\% \\
\hlinewd{1.2pt}
\rowcolor[gray]{.8}[5.5pt][2.5pt] Global detection& 0(0,00\%) 
& 0(0,00\%) & 0(0,00\%) & 0(0,00\%) & 0(0,00\%) & 0,00\% \\
\hlinewd{1.2pt}
\end{tabular}
\end{footnotesize}
 \caption{PE Legitimate Samples detection rates (Com = Communication \& Exchange Applications, MM = Multimedia Applications, Off = Office Applications, Sec = Security Applications, SysU = System \& Utilities).}
\end{table*}

\begin{table*}
\centering
\begin{footnotesize}
\begin{tabular}{|l@{ }|l@{ }|l@{ }|l@{ }|l@{ }|l@{ }|l@{ }|r@{ }|}
\hlinewd{1.2pt} 
\rowcolor[gray]{0}\textcolor{white}{Behaviors} & \textcolor{white}{EmW} 
& \textcolor{white}{FdW} & \textcolor{white}{IrcW} & \textcolor{white}{P2PW} & \textcolor{white}{V} & \textcolor{white}{Gen} & \textcolor{white}{Global}\\
\hlinewd{1.2pt} 
Nb string ciphered & 1/51 & 0/4 & 1/26 & 0/30 & 3/61 & 10/30 & 15/202 \\
\hline
Nb body ciphered & 4/51 & 0/4 & 0/26 & 1/30 & 2/61 & 0/30 & 7/202 \\
\hline
String encryption detection & 1(100\%) & 0 & 0 & 0(0\%) & 2(66,67\%) & 10(100\%) & 86,67\%\\
\hlinewd{1.2pt}
\rowcolor[gray]{.8}[5.5pt][2.5pt] Duplication & 43(84,31\%) 
& 4(100\%) & 20(76,96\%) & 22(73,33\%) & 44(72,13\%) & 30(100\%) & 80,70\% \\ 
\hlinewd{1.2pt}
 By direct copy & 41(80,39\%) 
& 4(100\%) & 20(76,96\%) & 22(73,33\%) & 25(40,98\%) & 30(100\%) & 70,30\% \\
 By single read/write & 8(15,69\%) 
& 0(0\%) & 4(15,38\%) & 3(10\%) & 21(34,43\%) & 0(0\%) & 17,82\%\\
 By interleaved read/write & 1(1,96\%) 
& 0(0\%) & 0(0\%) & 0(0\%) & 8(13,11\%) & 0(0\%) & 4,46\%\\
\hlinewd{1.2pt}
\rowcolor[gray]{.8}[5.5pt][2.5pt] Propagation & 33(64,71\%) 
& 3(75\%) & 5(19,23\%) & 25(83,33\%) & 5(8,20\%) & 30(100\%) & 49,99\% \\ 
\hlinewd{1.2pt}
 By direct copy & 33(64,71\%) 
& 3(75\%) & 4(15,38\%) & 25(83,33\%) & 3(4,92\%) & 30(100\%) & 48,52\% \\
 By single read/write & 3(5,88\%) 
& 0(0\%) & 2(7,69\%) & 1(3,33\%) & 2(3,28\%) & 0(0\%) & 3,96\%\\
 By interleaved read/write & 0(0\%) 
& 0(0\%) & 0(0\%) & 0(0\%) & 0(0\%) & 0(0\%) & 0,00\%\\
\hlinewd{1.2pt}
\rowcolor[gray]{.8}[5.5pt][2.5pt] Residency & 32(62,75\%) 
& 4(100\%) & 20(76,92\%) & 18(60,00\%) & 20(32,79\%) & 30(100\%) & 61,39\%\\ 
\hlinewd{1.2pt}
\rowcolor[gray]{.8}[5.5pt][2.5pt] Overinfection test & 4(7,84\%) 
& 1(25\%) & 1(3,85\%) & 0(0\%) & 0(0\%) & 0(0\%) & 2,97\%\\ 
\hlinewd{1.2pt}
by conditional 1 & 4(7,84\%) 
& 1(25\%) & 1(3,85\%) & 0(0\%) & 0(0\%) & 0(0\%) & 2,97\%\\ 
by inverse conditional 2 & 0(0\%) 
& 0(0\%) & 0(0\%) & 0(0\%) & 0(0\%) & 0(0\%) & 0,00\%\\ 
\hlinewd{1.2pt}
\rowcolor[gray]{.8}[5.5pt][2.5pt] Global detection & 46(90,20\%) 
& 4(100\%) & 25(96,15\%) & 27(90,00\%) & 50(81,97\%) & 30(100\%) & 90,09\%\\ 
\hlinewd{1.2pt}
\end{tabular}
\end{footnotesize}
 \caption{VBScript Malware detection rates (EmW = Email Worms, FdW = Flash Drive Worms, IrcW = IRC Worms, P2PW = Peer-to-peer Worms, V = Virii, Gen = Variants from malware generators).}
\end{table*}

\begin{table*}
\centering
\begin{footnotesize}
\begin{tabular}{|l@{ }|l@{ }|l@{ }|l@{ }|l@{ }|l@{ }|l@{ }|r@{ }|}
\hlinewd{1.2pt} 
\rowcolor[gray]{0}\textcolor{white}{Behaviors} & \textcolor{white}{EmM} 
& \textcolor{white}{InfC} & \textcolor{white}{Enc} & \textcolor{white}{DfE} & \textcolor{white}{MwC} & \textcolor{white}{RegR} & \textcolor{white}{Global}\\
\hlinewd{1.2pt} 
Nb string ciphered & 0/14 & 0/10 & 0/4 & 0/16 & 0/8 & 0/8 & 0/50\\
\hline
Nb body ciphered & 0/14 & 0/10 & 0/4 & 0/16 & 0/8 & 0/8 & 0/50\\
\hline
String encryption detection & 0 & 0 & 0 & 0 & 0 & 0 & 0\%\\
\hlinewd{1.2pt}
\rowcolor[gray]{.8}[5.5pt][2.5pt] Duplication & 0(0\%) 
& 0(0\%) & 0(0\%) & 0(0\%) & 0(0\%) & 0(0\%) & 0\%\\ 
\hlinewd{1.2pt}
 By direct transfer & 0(0\%) 
& 0(0\%) & 0(0\%) & 0(0\%) & 0(0\%) & 0(0\%)  & 0\%\\ 
 By single read/write & 0(0\%) 
& 0(0\%) & 0(0\%) & 0(0\%) & 0(0\%) & 0(0\%)  & 0\%\\ 
 By interleaved read/write & 0(0\%) 
& 0(0\%) & 0(0\%) & 0(0\%) & 0(0\%) & 0(0\%)  & 0\%\\ 
\hlinewd{1.2pt}
\rowcolor[gray]{.8}[5.5pt][2.5pt] Propagation & 0(0\%) 
& 0(0\%) & 0(0\%) & 0(0\%) & 0(0\%) & 0(0\%)  & 0\%\\ 
\hlinewd{1.2pt}
 By direct transfer & 0(0\%) 
& 0(0\%) & 0(0\%) & 0(0\%) & 0(0\%) & 0(0\%)  & 0\%\\ 
 By single read/write & 0(0\%) 
& 0(0\%) & 0(0\%) & 0(0\%) & 0(0\%) & 0(0\%)  & 0\%\\ 
 By interleaved read/write & 0(0\%) 
& 0(0\%) & 0(0\%) & 0(0\%) & 0(0\%) & 0(0\%)  & 0\%\\
\hlinewd{1.2pt}
\rowcolor[gray]{.8}[5.5pt][2.5pt] Residency & 0(0\%) 
& 0(0\%) & 0(0\%) & 0(0\%) & 1(12,50\%) & 0(0\%)  & 1,67\%\\ 
\hlinewd{1.2pt}
\rowcolor[gray]{.8}[5.5pt][2.5pt] Overinfection test & 0(0\%) 
& 0(0\%) & 0(0\%) & 0(0\%) & 0(0\%) & 0(0\%)  & 0\%\\ 
\hlinewd{1.2pt}
by conditional 1 & 0(0\%) 
& 0(0\%) & 0(0\%) & 0(0\%) & 0(0\%) & 0(0\%)  & 0\%\\ 
by inverse conditional 2 & 0(0\%) 
& 0(0\%) & 0(0\%) & 0(0\%) & 0(0\%) & 0(0\%)  & 0\%\\ 
\hlinewd{1.2pt}
\rowcolor[gray]{.8}[5.5pt][2.5pt] Global detection & 0(0\%) 
& 0(0\%) & 0(0\%) & 0(0\%) & 1(12,5\%) & 0(0\%)  & 1,67\%\\ 
\hlinewd{1.2pt}
\end{tabular}
\end{footnotesize}
\caption{VBScript Legitimate Samples detection rates (EmM = Email Managers, InfC = Informattion Collectors, Enc = Encoders, DfE = Disk and File Explorers, MwC = Malware Cleaners, RegR = Registry Repairs).}
\end{table*}

\subsection{Limitation of the collect mechanisms}
	As said in the previous section, a significant part of the false negatives is not due neither to the abstraction process nor the detection algorithm, but to the coverage of the collection mechanisms. Several limitations existing in collection mechanisms can directly explain the missed detections of certain behaviors. However, since our approach is based on separate layers, collection and abstraction can be improved for a given platform or language without modifying the upper detection layer.\\

\subsubsection{Dynamic analysis (PE traces)}
\indent Due to the dynamic nature of the collection, the first reason for detection failure is a problem related to the configuration of the simulated environment. The simulation must seem as real as possible in order to satisfy the execution conditions of the malware, in particular for triggered actions. \\
This problem can reside in the software configuration. Considering PE Virii, 64,6\% of the tested samples (53/82) did not execute properly in the simulated environment: invalid PE files, access violations or raised exceptions. Most file infector virii were released before the year 2000, explaining that the recent configuration of the simulation environment may not support their execution. Exceptions can also be used as anti-debug techniques crafted to hinder dynamic analysis.\\
The configuration problem can also reside in the simulated network. Considering Mail Worms, their propagation is conditioned by the network configuration. 75\% of the Mail Worms (45/60) did not show any SMTP activity simply because the required server was not reachable. Likewise, Net Worms propagate through vulnerabilities only if a vulnerable target is reachable. The absence of potential targets explains that 93,33\% of the worms did not show any propagation (14/15). All actions conditioned by the configuration of the simulated environment are difficult to observe: a potential solution could be forced branching.\\

\indent Beyond the configuration problem, the level of the collection can also explain the detection failure. With a low level collection mechanism, the visibility scope over the performed actions and the data flow is increased. All flow-sensitive behaviors such as duplication can be missed because of a breakdown in the data flow. Such breakdowns can find their origin in non monitored system calls but above all in the intervention of intermediate buffers where all operations are executed in memory. These buffers are often used in code mutations for example (polymorphism, metamorphism). Considering once again Virii, 12,20\% additional samples (10/82) were missed because of a data flow breakdown. The problem is identical with the mail propagation: 8,33\% of the propagations (5/60) were missed for Mail Worms because of an intermediate buffer intervening in the Base 64 encoding. These problems do not come from the grammatical signature of the behavior but from NtTrace which does not capture processor instructions. More complete collection tools either collecting instructions \cite{CA08} or deploying tainting techniques \cite{Anu,Arg} could avoid these breakdowns in the data flow.\\

\subsubsection{Static analysis (VB scripts)}
		Considering VB Scripts, the interpreted nature of the language implies a different context where the whole code is made available. Therefore, using static analysis, branching exploration becomes feasible and the whole data flow becomes observable. The VB Script analyzer implements these features, compensating for the drawbacks of NtTrace and eventually resulting in better detection rates.\\
\indent On the other hand, contrary to the relatively restricted number of system calls, the VB Script language offers numerous services to monitor. A same operation can be achieved using different managers or interfacing with different Microsoft applications. Considering the actual version of the analyzer, additional features could be monitored for a greater coverage: accesses to Messenger services or the support of the \texttt{"Windows Management Instrumentation (WMI)"} which would require parsing database requests. For example, listing connected drives for propagation is currently supported by the analyzer but this same list could be recovered using \texttt{WMI} by querying the \texttt{LogicalDisk} entries from the \texttt{"Win32\_ComputerSystem"} object.\\
\indent Moreover, like any other static analysis, the script analysis can be hindered by encryption and obfuscation techniques. The current version of the analyzer can partially handle these techniques as described in the section 5.B. Globally, static analysis is easier to consider with scripts because no prior disassembly is required and several security locks ease the analysis: no dynamic code rewriting, no dynamically resolved jumps. However, recent works have shown that inserting an intermediate interpretation layer could reintroduce all obfuscation techniques possible in low level languages (C, assembly) \cite{MR08}.\\

\subsection{Behavior relevance}
	The previous section deals with problems related to the collection mechanisms, but the behavioral model must be assessed itself. The relevance of each behavior must be individually assessed by checking the coverage of its grammatical model. Once the relevances determined, it becomes possible to extrapolate possible correlations between the different behaviors.\\
\indent Some behaviors such as duplication, propagation and residency are obviously characteristic to malware. Duplication and propagation are enough discriminating for detection. The only one of these behaviors likely to occur in legitimate programs is residency, during a program installation for example. The behavioral model should thus be refined in future works, using additional constraints on the value written to the booting object: the value should refer to the program itself or to one of its duplicated versions. This modification could help avoiding the remaining false positives observed.\\
\indent On the other hand, the model of the overinfection test does not seem completely relevant. The problem comes from a description too much restraint, which limits its detection. In particular, the conditional structure intervening in the model can not be detected in system call traces. A generalization of the model would increase its detection but the risk of confusion with error handling in legitimate programs would heavily increase. For future works, it would be interesting to test a new description of the overinfection test as well as additional behaviors.

\subsection{Performance}

\begin{center}
\begin{footnotesize}
\begin{tabular}{|l|l|l|}
\hlinewd{1.2pt}
\cellcolor[gray]{0}  \textcolor{white}{NtTrace} & \multicolumn{2}{|l|}{\cellcolor[gray]{0.8} Data reduction from PE traces to logs}\\
\hline
\cellcolor[gray]{0}  \textcolor{white}{Analyzer} & Total size: 351,32Mo & Average: 1,32Mo/Trace\\
\cellcolor[gray]{0}  & Reduced logs: 11,85Mo & Reduction ratio: 29\\
\hlinewd{1.2pt}
\cellcolor[gray]{0} & \multicolumn{2}{|l|}{\cellcolor[gray]{0.8} Execution speed}\\
\hline
\cellcolor[gray]{0} & Single core M 1,4GHz & Dual core 2,6GHz\\
\hline
\cellcolor[gray]{0} & 1,48 s/trace & 0,34 s/trace\\
\hlinewd{1.2pt}
\cellcolor[gray]{0}  \textcolor{white}{VB Script} & \multicolumn{2}{|l|}{\cellcolor[gray]{0.8} Data reduction from VB scripts to logs}\\
\hline
\cellcolor[gray]{0}  \textcolor{white}{Analyzer} & Total size: 1842Ko & Average: 7Ko/Script\\
\cellcolor[gray]{0}  & Reduced logs: 298Ko & Reduction ratio: 6\\
\hlinewd{1.2pt}
\cellcolor[gray]{0} & \multicolumn{2}{|l|}{\cellcolor[gray]{0.8} Execution speed}\\
\hline
\cellcolor[gray]{0} & Single core M 1,4GHz & Dual core 2,6GHz\\
\hline
\cellcolor[gray]{0} & 0,042 s/script & 0,016 s/script\\
\cellcolor[gray]{0} & +0,50 s/ciphered line & +0,21 s/cipered line\\
\hlinewd{1.2pt}
\cellcolor[gray]{0}  \textcolor{white}{Detection} & \multicolumn{2}{|l|}{\cellcolor[gray]{0.8} Execution speed}\\
\hline
\cellcolor[gray]{0}  \textcolor{white}{Automata} & Single core M 1,4GHz & Dual core 2,6GHz\\
\hline
\cellcolor[gray]{0} & NT: 0,44 s/log & NT: 0,14 s/log\\
\cellcolor[gray]{0} & VBS: 0,002 s/log & VBS: $<$0,001 s/log\\
\hline
\end{tabular}
\end{footnotesize}
\end{center}

\indent The table above provides the measured performances for the different components of the prototype. With regards to the abstraction layer, the analysis of PE script seems the most time consuming task. This is not surprising since the analyzer uses lots of string comparisons which could be partially avoided by replacing the off-line analysis by real time collection and translation. By hooking directly the system calls, the translation becomes immediate and does not require cumbersome comparisons. On the other hand, the Visual basic script analyzer seems lighter and offers satisfying performances. Once optimized, it could be deployed on mail servers to analyze joint pieces for example.\\
\indent With regards to the detection automata, the performances are also satisfying compared with the worst case complexity defined in Proposition 1. The detection speed remains far below the order of a half second in more than 90\% of the cases; the remaining 10\% cases were all malware. The automata implementation has also revealed that the maximum required space for the syntactic and semantic stacks was very low: 7 elements and 3 elements are the respective maximal sizes reached by the prototype for the syntactic and semantic stacks ($2s<10$ in Proposition 1.). In addition to speed, the number of raised ambiguities has also been measured. If we denote $n_e$ the number of events and $n_a$ the number of ambiguity. In a worst case scenario, we would have $n_a=2^{n_e}$. By experience we have:\\
$n_a<<2^{n_e}$\\
$n_a<<n_e^2$\\
$n_a\approx \alpha n_e$\\
\textbf{This approximation provides a new practical complexity in $\vartheta(k\alpha(\frac{n^2+n}{2}))$ which is more worth considering.} Moreover, this algorithm can easily be parallelized for optimization in the new multi-core architectures. The Figure 12 and 13 provide graphs of the collected $\alpha$ ratios. \textbf{From these graphs, it can be observed that above a certain threshold, an important ratio of ambiguity $\alpha$ is already a sign of malicious activity}. 

\begin{figure}[!ht]
  \centering
  \includegraphics*[scale=0.32]{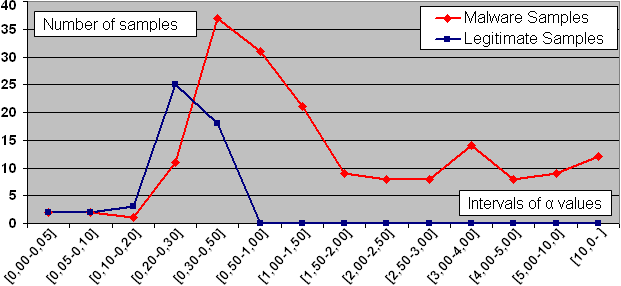} 
  \label{fig:peamb}
%  \usefont{OT1}{cmr}{b}{n}
  \caption{PE Ambiguity Ratio. Abscissa: intervals of $\alpha$ values / Ordinate: number of samples in the interval.}
\end{figure}
\begin{figure}[!ht]
  \centering
  \includegraphics*[scale=0.32]{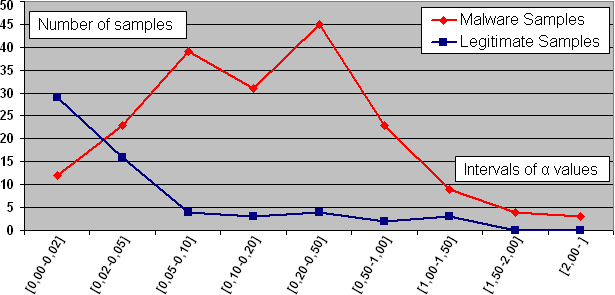} 
  \label{fig:vbsamb}
%  \usefont{OT1}{cmr}{b}{n}
  \caption{VBS Ambiguity Ratio. Abscissa: intervals of $\alpha$ values / Ordinate: number of samples in the interval.}
\end{figure}

\section{Conclusions}	
	Detection by attribute automata provides a good detection coverage of malware using known techniques with 51\% of detected PE malware and 89\% of Visual Basic Scripts malware. The grammatical approach offers a synthetic and understandable vision of malicious behaviors. Indeed, only four generic, human-readable, behavioral descriptions have resulted in these significant detection rates. Unknown malware using variations from these known techniques should remain detected thanks to the abstraction process. In case of innovative techniques, this approach eases the update process. Thanks to the decoupled layers for abstraction and detection, updates can be independently applied at two levels: in the grammatical descriptions in case of new generic procedures (the less frequent), or in the abstraction components in case of new vulnerable objects or API (the most frequent).\\
\indent Up until now, the generation of the behavioral descriptions is still manual but the process could be combined with the identification of malicious behaviors by differential analysis proposed by Christodorescu, Jha and Kruegel \cite{CJK07}.\\
\indent The experimentations have also stressed the importance of the collection mechanism in the detection process. Collection mechanisms are already an active research field and future work can be testing more adapted collection tools deploying tainting.\\

\thanks{\textbf{Acknowledgement:} This work has been partially supported by the European Commissions through project FP7-ICT-216026-WOMBAT funded by the 7th framework program. The opinions expressed in this paper are those of the authors and do not necessarily reflect the views of the European Commission.}

\bibliographystyle{IEEEtran}
\bibliography{IEEEabrv,references}

\end{document}